\documentclass[aps,pre,preprint,showpacs,superscriptaddress,nofootinbib]{revtex4}
\usepackage{amsmath}
\usepackage{epsfig}

\newcommand{\be}{\begin{equation}}
\newcommand{\ee}{\end{equation}}
\newcommand{\bea}{\begin{eqnarray}}
\newcommand{\eea}{\end{eqnarray}}

\newcommand{\pr}{\prime}
\newcommand{\sn}{{\rm sn}}

\newcommand{\dn}{{\rm dn}}
\newcommand{\cn}{{\rm cn}}
\newcommand{\sech}{{\rm sech}}

\begin{document}

\bibliographystyle{revtex}
\title{Dynamic and Static Excitations of a Classical 
Discrete Anisotropic Heisenberg Ferromagnetic Spin Chain} 

\author{M. Lakshmanan} 
\address{Centre for Nonlinear Dynamics, Department of Physics, 
Bharathidasan University, Tiruchirapalli 620 024, India} 

\author{Avadh Saxena}
\address{Theoretical Division and Center for Nonlinear Studies, Los
Alamos National Lab, Los Alamos, NM 87545, USA}


\begin{abstract}
Using Jacobi elliptic function addition formulas and summation 
identities we obtain several static and moving periodic soliton 
solutions of a classical anisotropic, discrete Heisenberg spin 
chain with and without an external magnetic field.  We predict the 
dispersion relations of these nonlinear excitations and contrast 
them with that of magnons and relate these findings to the 
materials realized by a discrete spin chain.  As limiting cases, 
we discuss different forms of domain wall structures and their 
properties.  
\end{abstract}
\maketitle
\section{Introduction} 
Establishing integrability and obtaining exact solutions of discrete 
nonlinear physical systems are important issues of current interest. 
Starting with the integrable discrete model of Ablowitz and Ladik 
\cite{AL}, for several other discrete nonlinear evolution equations 
exact elliptic function and soliton solutions have been obtained in 
recent years \cite{scott}. These include certain discrete versions 
of nonlinear Schr\"odinger (NLS) equation \cite{dnls,dnls2}, $\phi^4$ 
equation \cite{phi4,phi4t}, derivative NLS equation \cite{dnls1}, 
coupled $\phi^4$ equation \cite{ks1}, coupled asymmetric double well 
and coupled $\phi^6$ equation \cite{ks2}, complex modified Korteweg-de 
Vries equation \cite{chaos}, etc., where effective use of summation 
relations of Jacobian elliptic functions was made and periodic and 
solitary wave solutions of moving and static types obtained. 

In this connection, a physically important discrete nonlinear 
dynamical system which has been of considerable interest in diverse 
areas of physics for a long time is the anisotropic Heisenberg 
ferromagnetic (and antiferromagnetic) spin system with or without 
an external magnetic field.  It has been studied for various aspects 
in magnetism, condensed matter physics/materials science, statistical 
physics, nonlinear dynamics, etc. both from classical and quantum points 
of view \cite{mattis}.  For example, the one-dimensional quantum spin-1/2 
XYZ chain has been shown to be an exactly solvable system either 
through Bethe ansatz procedure or through quantum inverse scattering 
method \cite{faddeevbook,baxter} and the eigenvalue spectrum and 
eigenfunctions have been obtained.  For large value of spins, however, 
a classical/quasiclassical description has been known to be an 
adequate description so that spins can be treated as unit vectors 
and classical equations of motion for the spin vectors can be obtained 
as limiting forms of the quantum equation of motion or as dynamical 
equations derived from postulated spin Poisson bracket relations 
\cite{discrete1}.  Another area of considerable physical interest in 
which such classical anisotropic spin systems have been studied in 
the presence of Gilbert damping is the microscopic behavior of spin 
waves in magnetic bodies of arbitrary shape \cite{Rivkin} and the study 
of spin-torque effect in  ferromagnetic layers with spin currents 
\cite{ml} on spin waves and domain walls. Recently it has also been 
pointed out that discrete breathers can exist in anisotropic spin
chains with additional onsite anisotropy \cite{Zolotaryuk}.  In any 
case, the resultant equations of motion describe an extremely 
interesting class of discrete nonlinear dynamical systems and 
exploration of the underlying dynamical properties is of interest 
both from theoretical and applied physics points of view. 
   
The one-dimensional Heisenberg ferromagnetic spin system with nearest 
neighbor exchange interaction has been shown to possess several completely 
integrable soliton bearing systems in its {\it continuum limit}: (i) the 
pure isotropic case \cite{laksh,takhtajan}, (ii) the uniaxial anisotropic 
case \cite{BNS}, and (iii) the biaxial anisotropic case \cite{sklyanin}.  
These systems also have a strong connection with the nonlinear Schr\"odinger 
equation \cite{Laksh}.  However, till date no exactly integrable 
{\it discrete} dynamical Heisenberg spin system has been identified in 
the literature, although a variant of the system, namely the Ishimori spin 
chain, is known to be completely integrable \cite{ishimori}.  It is 
generally expected that the discrete anisotropic Heisenberg spin chain 
is a nonintegrable nonlinear dynamical system.  Yet, as we show in this 
article, a number of interesting exact periodic and stationary 
structures, including domain wall type structures, for the fully 
anisotropic system (XYZ case as well as the limitng XYY and planar XY 
cases) can be obtained and their properties analyzed using standard 
techniques.  In fact, Roberts and Thompson \cite{roberts} and Granovskii 
and Zhedanov in a series of papers \cite{discrete2,jetp,granov}, have 
obtained special classes of solutions for the anisotropic spin system.  
In particular, the latter authors have shown that the time-independent  
case of the XYZ anisotropic spin chain is an integrable map by 
relating it to a Neumann type discrete system \cite{jetp}, see also 
Ref. \cite{veselov}.  

In this paper, by parameterizing the unit spin vector in terms of 
the basic Lam\'e polynomials of lower order \cite{lame} or their 
derivatives and by a judicious use of various addition theorems and 
summation relations obeyed by Jacobi elliptic functions \cite{byrd} 
we point out that several classes of explicit dynamical and static 
structures can be obtained. In the limiting cases we obtain linear spin 
wave solutions and different nonlinear domain wall type solutions in a 
natural way.  We study the physical implications of these solutions like 
the energy spectrum, effect of discreteness such as the Peierls-Nabarro 
barrier \cite{nabarro,PN,soboleva}, linear stability and so on. 

The plan of the paper is as follows.  In Sec. II, we introduce the 
dynamical equations of motion and introduce certain natural 
parametrizations of the unit spin vector.  In Sec. III, we obtain 
two classes of periodic solutions, investigate the associated 
dispersion relations and energy expressions and indicate a 
semiclassical quantization of these solutions.  In Sec. IV we 
report various classes of static solutions for the XYZ, XYY and XY 
planar models.  In Sec. V, we obtain the total energy expressions    
associated with the various static solutions and discuss the effect 
of discreteness including the Peierls-Nabarro potential barrier.  In 
Sec. VI, the isotropic case is considered, while in Sec. VII the linear 
stability of both time periodic and static solutions is investigated. 
Then in Sec. VIII, we present some explicit time dependent solutions for the 
case when the onsite anisotropy or an external magnetic field is introduced.  
Finally, in Sec. IX we summarize our results.  In the Appendix A we include 
some of the relevant addition theorems and summation relations obeyed by the 
Jacobi elliptic functions required for our analysis, while in Appendix B 
some details on semiclassical quantization are given.   
 
 
\section{The Heisenberg Anisotropic Spin Chain} 

\subsection{Equation of Motion} 

We consider a one dimensional anisotropic Heisenberg ferromagnetic 
spin chain with the spin components $\vec{S_n}=(S_n^x,S_n^y,S_n^z)$, 
satisfying the constraint of unit length  
\be 
(S_n^x)^2+(S_n^y)^2+(S_n^z)^2=1 , 
\ee 
modeled by the Hamiltonian 
\be 
H=-\sum_{\{n\}}(AS_n^xS_{n+1}^x+BS_n^yS_{n+1}^y+CS_n^zS_{n+1}^z) 
-D\sum_n(S_n^z)^2-\vec{\cal H}\cdot\sum_n{\vec{S}}_n , 
\ee 
where the sum is over the nearest neighbors, $A$, $B$ and $C$ are 
the (exchange) anisotropy parameters, $D$ is the onsite anisotropy parameter
 and $\vec{\cal H}=({\cal H},0,0)$ is
the external magnetic field along the $x$ direction (for convenience). 
For the XYZ model, $A\ne B\ne C, D=0$ and for the XY model $C=0$ and $D=0$.  
Using the spin Poisson bracket relation \cite{discrete1} 
\be 
\{S_i^{\alpha},S_j^{\beta}\}_{PB} = \delta_{ij}\epsilon_{\alpha\beta \gamma} 
S_j^{\gamma} , ~~~ \alpha,~\beta,~\gamma=1,2,3 ,  
\ee 
where $\delta_{ij}$ is the Kronecker delta and $\epsilon_{\alpha\beta\gamma}$ 
is the Levi-Civita tensor, for any two functions ${\cal A}$ and 
${\cal B}$ of spins one has 
\be 
\{{\cal A},{\cal B}\}_{PB} =\sum_{\alpha,\beta,\gamma}
\sum_{i=1}^{N}\epsilon_{\alpha\beta \gamma}\frac{\partial 
{\cal A}}{\partial S_i^{\alpha}}\frac{\partial {\cal B}}{\partial 
S_i^{\beta}} S_i^{\gamma} , 
\ee
and the equation of motion becomes  
\begin{eqnarray} 
\frac{d\vec{S}_n}{dt}=\vec{S_n}\times[A(S_{n+1}^x+S_{n-1}^x)\vec{i} 
+B(S_{n+1}^y+S_{n-1}^y)\vec{j} + C(S_{n+1}^z+S_{n-1}^z)\vec{k}
\nonumber \\ +2 D S_n^z\vec{k}] 
+ \vec{S}_n\times\vec{\cal H},\;\;\;\; 
n=1,2,...,N,
\end{eqnarray}  
where $\vec{i},\,\vec{j},\,\vec{k}$ form a triad of Cartesian unit vectors.
Explicitly, in component form the above equation reads  
\be \frac{d{S_n^x}}{dt}=CS_n^y(S_{n+1}^z+S_{n-1}^z) 
-BS_n^z(S_{n+1}^y+S_{n-1}^y)-2DS_n^yS_n^z ,  
\ee

\be \frac{d{S_n^y}}{dt}=AS_n^z(S_{n+1}^x+S_{n-1}^x) 
-CS_n^x(S_{n+1}^z+S_{n-1}^z)+2DS_n^xS_n^z + {\cal H}S_n^z,  
\ee

\be \frac{d{S_n^z}}{dt}=BS_n^x(S_{n+1}^y+S_{n-1}^y) 
-AS_n^y(S_{n+1}^x+S_{n-1}^x) - {\cal H}S_n^y.   
\ee
Equations (5) or (6)-(8) can also be obtained as the limiting case 
of the corresponding quantum dynamical equation of motion for the 
spin operators when $\hbar\rightarrow0$ or $S\rightarrow\infty$. In 
either case, the dynamics is obtained by solving the initial value 
problem of the system of coupled nonlinear ordinary differential 
equations (5) or (6)-(8) along with the constraint (1) on the spin 
vectors, subject to appropriate boundary conditions like $S_n 
\rightarrow_{n\rightarrow\infty}(\pm1,0,0)$ or $S_n\rightarrow_{n 
\rightarrow\infty}(0,0,\pm1)$.  However, it appears that the system 
of differential equations (6)-(8) is in general nonintegrable.  Even 
then one can obtain several special classes of solutions of physical 
interest by making use of the properties of (Jacobian) elliptic 
functions and parametrizing the spin vector to satisfy the unit length 
condition (1).  As noted in the Introduction, some of these solutions 
were reported earlier by Roberts and Thompson \cite{roberts}, and by 
Granovskii and Zhedanov \cite{granov}, which are to be discussed in 
the following sections; however, as we point out in this paper a much 
larger class of explicit exact solutions can be found in a rather 
transparent manner through appropriate parametrizations of the spin 
vectors.  

Before dwelling upon the discrete chain, it is also of interest to 
note as pointed out in the Introduction that the long wavelength/low 
temperature continuum limit of Eqs. (6)-(8), when the lattice 
parameter $a\rightarrow0$, takes the form (in the $D=0$ limit)
\be 
\frac{\partial\vec{S}(x,t)}{\partial t} = \vec{S}\times\vec{J}~  
\frac{\partial^2\vec{S}}{\partial x^2} + \vec{S}\times\vec{\cal H} , 
\ee 
where $\vec{\cal H}=({\cal H},0,0)$, $\vec{J}\vec{S}=AS_x\vec{i}+BS_y\vec{j}
+CS_z\vec{k}$, 
$S_x^2+S_y^2+S_z^2=1$ and $A$, $B$, and $C$ are the anisotropy parameters. 
The isotropic case $A=B=C$ is a completely integrable soliton system 
and is equivalent to a nonlinear Schr\"odinger equation in a geometrical 
\cite{laksh} and gauge equivalence sense \cite{takhtajan}.  So are  
the uniaxial anisotropic spin chain $(A=B\ne C)$ in the presence of 
a longitudinal magnetic field \cite{BNS} and the biaxial anisotropic 
spin chain without the magnetic field \cite{sklyanin} integrable soliton 
systems.  In spite of the existence of these integrable continuum spin 
systems, the discrete chain (5) remains as a rather difficult problem 
to analyze. 

\subsection{The Parametrization of the Unit Spin Vector} 

One way to proceed with the analysis is to start with an appropriate 
parametrization of the unit sphere of spin given by Eq. (1).  Obviously   
natural parametrizations are in terms of elliptic functions.  For this  
purpose one can start with the eigenfunctions of the Lam\'e equation 
\be 
\frac{d^2\psi(u)}{du^2} + [E-n(n+1)k^2\sn^2(u,k)]\psi(u) = 0 
\ee 
for positive integer $n$, which are given in terms of Lam\'e 
polynomials.  The lowest order ($n=1$) polynomials are \cite{lame} 
\be 
\psi_{11}\propto \sn(u,k), ~~~ \psi_{12}\propto \cn(u,k) , ~~~ 
\psi_{13}\propto \dn(u,k) , 
\ee 
while the next order ones are ($n=2$) 
\be 
\psi_{21}\propto \sn(u,k)\cn(u,k), ~~~ \psi_{22}\propto \cn(u,k)\dn(u,k), 
~~~ \psi_{23}\propto \sn(u,k)\dn(u,k) , 
\ee 
and so on.  Here $\sn(u,k)$, $\cn(u,k)$ and $\dn(u,k)$ are the 
standard Jacobian elliptic functions \cite{byrd} characterized by the 
modulus parameter $k$ [see also Appendix A for the relevant properties 
of the Jacobian elliptic functions].  Consequently we can choose, for 
example, an appropriate set of parametrization for the unit spin 
vectors as 
\be 
S_n^x=\alpha~\sn(u,k) , ~~~ S_n^y=\beta~\cn(u,k), ~~~ S_n^z= 
\gamma~\dn(u,k) . 
\ee 
where $\alpha, \beta$ and $\gamma$ are constant parameters to be fixed.  
The requirement that condition (1) should be satisfied requires 
\be 
\alpha^2=1-\gamma^2+\gamma^2 k^2=1-\gamma^2 k'^2, ~~~\beta^2=1-\gamma^2 , 
\ee 
where $\gamma$ is a free parameter ($0\le\gamma\le1$) and $k$ is the 
modulus parameter and $k'=\sqrt{1-k^2}$ is the complementary modulus. One 
can easily check that a parametrization  $S_n^x=\alpha\sn(u,k) \cn(u,k)$, 
$S_n^y =\beta\cn(u,k)\dn(u,k)$ and $S_n^z=\gamma\sn(u,k) \dn(u,k)$ does not 
satisfy the condition (1) for any set of real values of $\alpha$, $\beta$ 
and $\gamma$. So one can proceed to higher order Lam\'e polynomials 
\cite{lame} for other possible parametrizations. 

One can even proceed with more general parametrizations 
in terms of two variables 
such as 
\be 
S_n^x=\cn(u,k_1), ~~~ S_n^y=\sn(u,k_1)\cn(v,k_2), ~~~ S_n^z=\sn(u,k_1) 
\sn(v,k_2) , 
\ee 
with two different moduli $k_1$ and $k_2$ or even more general forms such as 
\begin{eqnarray}  
S_n^x&=&\frac{\alpha\cn(u,k_1)}{1-\gamma\sn(u,k_1)\sn(v,k_2)}, ~~~ 
S_n^y=\frac{\alpha\sn(u,k_1)\sn(v,k_2)}{1-\gamma\sn(u,k_1)\sn(v,k_2)},
\nonumber \\  
S_n^z&=&\frac{\sn(u,k_1)\sn(v,k_2)-\gamma}{1-\gamma\sn(u,k_1)\sn(v,k_2)},     
 ~~~ 
\alpha=\sqrt{1-\gamma^2} ,  
\end{eqnarray}  
both of which satisfy condition (1).  We will also make use of these 
parametrizations in our analysis. 




\section{Moving Solutions: Anisotropic case ($D=0, \,\vec{\cal H}=0$)} 

We now look for time dependent moving solutions of Eqs. (6) - (8) 
when the onsite anisotropy and magnetic field are absent ($D=0$ and $\vec{\cal H}=0$) in the form 
(13) and (14) with the substitution $u=pn-\omega t+\delta$, so that 
\be 
S_n^x=\alpha~\sn(pn-\omega t+\delta,k), ~~~ S_n^y=\beta~\cn(pn-\omega 
t +\delta,k), ~~~ S_n^z= \gamma~\dn(pn-\omega t+\delta,k) , 
\ee 
along with the relations (14). Here $p$ and $\omega$ are the wave 
vector and angular frequency, respectively, which are to be fixed in 
conjunction with Eqs. (6) - (8) and $\delta$ is a phase constant.  On 
substituting the expressions (13) for the components of the spin 
vector $\vec{S}_n(t)$ into the Eqs. (6) - (8) and making use of the 
addition theorems for the Jacobian elliptic functions (see Appendix 
A), one requires the following conditions to be satisfied: 
\be 
-\omega\alpha=\frac{2\beta\gamma[C\dn(p,k)-B\cn(p,k)]}{1-k^2\sn^2(u,k) 
\sn^2(p,k)} ,  
\ee 
\be 
\omega\beta=\frac{2\alpha\gamma\dn(p,k)[A\cn(p,k)-C]}{1-k^2\sn^2(u,k)   
\sn^2(p,k)} ,  
\ee
\be 
\omega\gamma k^2=\frac{2\alpha\beta\cn(p,k)[B-A\dn(p,k)]}{1-k^2 
\sn^2(u,k)\sn^2(p,k)} ,  ~~~ u=pn-\omega t + \delta .    
\ee
In each of the above expressions, we note that the variable $u$ occurs explicitly 
on the right hand sides.  Consequently the $u$-dependent terms in the 
above relations can be  avoided iff $p$ and $\omega$ are chosen in one of
the following three ways:
\begin{enumerate}
\item{$\omega=0$, $p\ne0$,} 
\item{modulus parameter $k=0$ (linear spin wave solution, see below),}
\item{$p=4K(k)$, or $p=2K(k)$ (or integral multiples of the right hand 
sides), where $K(k)$ is the complete elliptic integral of the first kind 
\cite{byrd}.}
\end{enumerate}
In the above, case (1), $\omega=0$, corresponds to the existence of static
solutions.  These are discussed in detail in sec. IV below.

Case (2), $k=0$, corresponds to linear spin wave (or magnon) solutions.  
More details are given in the subsection IIIc below.

For case (3), note that $\sn(2K,k)=0 
=\sn(4K,k)$, $\cn(2K,k)=-1$, $\dn(2K,k)=1=\cn(4K,k)=\dn(4K,k)$.  Further, 
since $\frac{\pi}{2}\le K(k)<\infty$ as $0\le k<1$, $p$ in Eq. (\ref{eqn21}) 
is bounded below by $\pi$ or $2\pi$ as the case may be. Correspondingly, 
we can have two families of periodic solutions, each of which we will 
consider separately: 
\be p=4K(k) , ~~~~ or ~~~~ p=2K(k),
\label{eqn21}
\ee 

\subsection{Spatially homogeneous time-dependent solutions} 
For the choice $p=4K(k)$, the conditions (18) - (20) reduce to 
\be
\omega\alpha=-2\beta\gamma(C-B) , ~~~ \omega\beta=2\alpha\gamma(A-C) ,
~~~ \omega\gamma k^2=2\alpha\beta(B-A) . 
\ee
Solving (22), we obtain 
\be 
\omega=2\gamma\sqrt{(B-C)(A-C)} , ~~~ k^2=\frac{1-\gamma^2}{\gamma^2} 
\frac{(B-A)}{(A-C)} , ~~~ (B>A>C) ,  
\ee 
where $\gamma$ is a free parameter ($0\le\gamma\le1$).  The corresponding 
solutions are 
\be 
S_n^x=\sqrt{1-\gamma^2 k'^2}~\sn(4Kn-\omega t+\delta,k)=-\sqrt{1- 
\gamma^2 k'^2}~\sn(\omega t+\delta,k) , 
\ee 
\be 
S_n^y=\sqrt{1-\gamma^2}~\cn(4Kn-\omega t+\delta,k)=\sqrt{1-\gamma^2}~ 
\cn(\omega t+\delta,k) , 
\ee 
\be 
S_n^z=\gamma~\dn(4Kn-\omega t+\delta,k)=\gamma~\dn(\omega t+\delta,k) .
\ee 
The above spatially homogeneous and time periodic solution is nothing 
but the description of Poinsot's motion of a rigid body pointed out 
by Roberts and Thompson \cite{roberts}.  Each of the spins in the 
lattice precesses about one of the axes with nutation in the same manner. 
This is depicted schematically in Fig. 1, where all the spins precess parallel
to each other.  Note that the axes stand for the three spin components $S^x$,
$S^y$ and $S^z$.

\begin{figure}[!ht]
\begin{center}
\includegraphics[width=.9\linewidth]{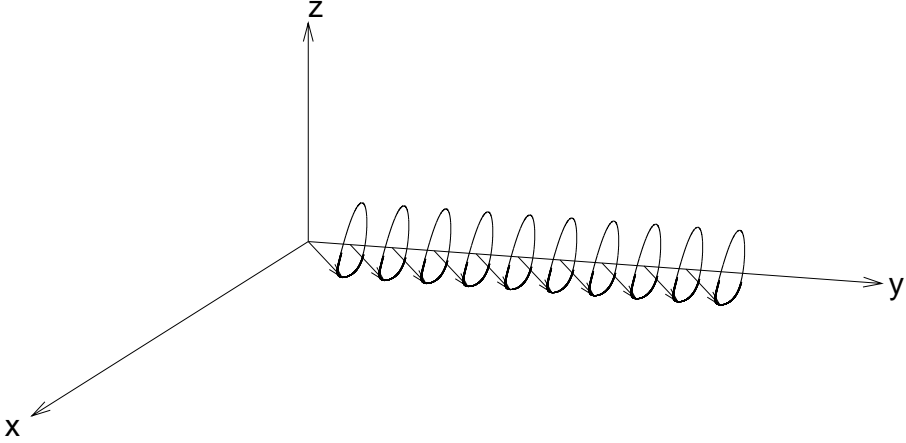}
\caption{Spatially homogeneous time-dependent solution for the spin vectors.}
\end{center}
\end{figure}

The total energy associated with the spin precession of $N$ nearest 
neighbor spins in a periodic lattice can be evaluated by substituting 
the solutions (24) - (26) into the energy expression (2) (with 
$D=0$ and $\vec{\cal H}=0$) and making use of the summation relations of the
Jacobi elliptic functions (Appendix A): 
\begin{eqnarray} 
E&=&-\sum_{n=1}^{N} [AS_n^x(S_{n+1}^x+S_{n-1}^x)+BS_n^y 
(S_{n+1}^y+S_{n-1}^y) + CS_n^z(S_{n+1}^z+S_{n-1}^z)] \nonumber \\ 
&=&-2N(B\beta^2+C\gamma^2)=-N[B+(C-B)\gamma^2]\nonumber \\
&=&-2N[B-(B-C)\gamma^2], \;~~~(B>C, ~~0\leq \gamma\leq1) 
\end{eqnarray} 
so that the energy per site becomes  
\be 
\epsilon=\frac{E}{N}=-2[B-(B-C)\gamma^2]. 
\ee 

Further, since each of these spins evolves identically, the spin chain 
may be treated to be equivalent to $N$ independent rigid bodies executing 
synchronous periodic motions of the form (24) - (26).  Consequently, each 
of the above spin motion can be quantized semi-classically using the 
Bohr-Sommerfeld quantization condition 
\be 
\oint p_idq_i = \left(n_i+\frac{1}{2}\right) h,\qquad n_i=0,1,2,..., ~~~ 
i=1,2,...,N  
\ee 
where the canonically conjugate variables 
\be 
p_n= S_n^z , ~~~ q_n=\arctan\left(\frac{S_n^y}{S_n^x}\right),~~~ n=1,2,...,N.
\ee 
Using the explicit forms for $S_n^x, S_n^y$ and $S_n^z$ given in (24) - (26),
$q_n$'s and $p_n$'s may be expressed in terms of elliptic functions (see 
Appendix B for details). 
Carrying out the integral over a cycle of period $4K$, one obtains the 
following transcendental equation for the quantization of the amplitude 
$\gamma$: 
\begin{eqnarray}
\frac{4}{\gamma}\sqrt{\frac{1-\gamma^2k'^2}{1-\gamma^2}}
\bigg[\Pi\bigg(\frac{-\gamma^2k^2}{(1-\gamma^2)},k\bigg)-(1-\gamma^2)K(k)\bigg]=\left(n_i+\frac{1}{2}\right)h,
\nonumber \\ 
n_i=0,1,2,...,~~~i=1,2,\ldots,N. 
\end{eqnarray} 
Here $\Pi(x,k)$ is the complete elliptic integral of the third kind 
\cite{byrd}.  Equation (31) is a transcendental equation in the amplitude 
parameter $\gamma$.  For each value of the quantum number $n_i$ (=0, 1, 2, ...), 
the solution $\gamma_{n_i}$ can be found by solving numerically the transcendental 
equation (31).  It may be noted that such a semiclassical quantization procedure 
by solving transcendental equations involving all the three complete elliptic 
integrals has been carried out successfully for isotropic anharmonic oscillators 
with two and three degrees of freedom \cite{athavan1} and for the two center 
Coulomb problem \cite{athavan2}.  Then using the resultant allowed set of values 
of the amplitude $\{\gamma_{n_i}\}$, $n_i$=0,1,2,... , $i=1,2,...,N$, in the 
classical energy expression per site (28), the corresponding quantized energy 
spectrum can be evaluated.  Consequently, the full spectrum associated with the 
solutions (24) - (26) of the total lattice can be evaluated by associating quantum 
numbers as $\{n_1, n_2,..., n_N\}$ with the full lattice.  Complete details will 
be published elsewhere. 

\subsection{Spatially oscillatory time periodic solutions}

Now taking the possibility $p=2K(k)$ in Eq. (18) for the wave vector, 
and using it in the conditions (19) and (20), we obtain the relations 
connecting the unknowns $\omega$, $\gamma$ and $k$ as 
\be
\omega\alpha=-2\beta\gamma(B+C) , ~~~ \omega\beta=-2\alpha\gamma
(A+C) , ~~~ \omega\gamma k^2=-2\alpha\beta(B-A) .
\ee
Solving these equations, we obtain 
\be
\omega=2\gamma\sqrt{(A+C)(B+C)} ,~~~ k^2= \frac{1-\gamma^2}{\gamma^2} 
\left(\frac{B-A}{A+C}\right) , ~~~ p=2K(k) .
\ee
The corresponding spatially alternating time periodic solutions are 
\be 
S_n^x=\sqrt{1-\gamma^2 k'^2}~\sn(2Kn-\omega t+\delta,k)=(-1)^{n+1} 
\sqrt{1-\gamma^2k'^2}~\sn(\omega t+\delta,k) , 
\ee 
\be 
S_n^y=\sqrt{1-\gamma^2}~\cn(2Kn-\omega t+\delta,k)=(-1)^n\sqrt{1- 
\gamma^2}~\cn(\omega t+\delta,k) , 
\ee 
\be 
S_n^z=\gamma~\dn(2Kn-\omega t+\delta,k)=\gamma~\dn(\omega t+\delta,k) . 
\ee
The solution (34) - (36) is depicted schematically in Fig. 2.  
Note that the $x$ and $y$ components of the alternate spins flip and next
nearest neighbors evolve in parallel.  Also, these solutions have no 
counterpart in the continuum limit of the lattice.  

\begin{figure}[!ht]
\begin{center}
\includegraphics[width=.9\linewidth]{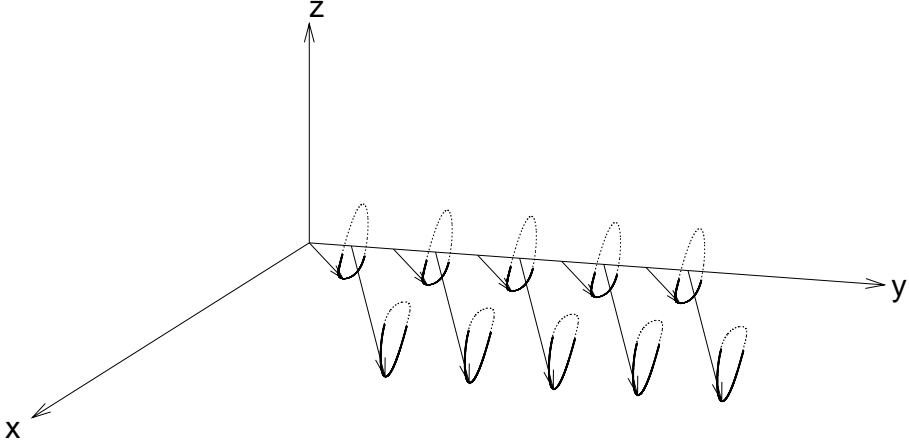}
\caption{Spatially oscillatory time periodic solution.}
\end{center}
\end{figure}

Again from the energy expression in Eq. (2), using the above solution 
(34) - (36), we obtain the total energy of the system for a periodic 
lattice of $N$ spins (with $N$ even) 
\be 
E=N[B-(B+C)\gamma^2] . 
\ee 
Correspondingly the energy per lattice site is 
\be 
\epsilon=[B-(B+C)\gamma^2] , 
\ee 
which is greater than the energy of the uniform periodic solution, 
see Eq. (28).  Consequently, the present solution constitutes an 
excited state of the system. 

Since every other spin evolves identically, a lattice of $N$ spins 
($N$ even) may be split into two sublattices of $N/2$ spins, and each 
member of the sublattice evolves identically.  Consequently, the 
semiclassical quantization condition may be given separately for each 
member of the two sublattices: 
\bea 
\oint p_{1,i} dq_{1,i}=\left(n_{1,i}+\frac{1}{2}\right)h, ~~~
n_{1,i}=0,1,2,\ldots, ~~~i=1,2,\ldots \frac{N}{2}\nonumber\\
\oint p_{2,i} dq_{2,i}=\left(n_{2,i}+\frac{1}{2}\right)h, ~~~
n_{2,i}=0,1,2,\ldots, ~~~i=1,2,\ldots \frac{N}{2} 
\eea 
and the energy expression may be correspondingly quantized with $q_i$ and 
$p_i$ chosen in the form (30). From (34) - (36) one can choose for the 
first sublattice the spin solutions corresponding to $n$ odd and for the 
second sublattice corresponding to $n$ even in the solution (34) - (36). 
Then evaluating the integrals (39), which result in expressions essentially 
of the form (31), the energy expression (37) can be quantized, as discussed 
earlier. 

\subsection{The linear and nonlinear magnon solutions and dispersion relations} 

In the uniaxial anisotropic case $A=B<C$, from the expression 
(33) for $k^2$, we find that $k=0$.  Consequently, we have the 
magnon solution. Considering the case (2), $k^2=0$ in Eqs (18) - (20), we find that
here one has the standard dispersion relation for the uniaxial anisotropic case
$(A=B<C)$ 
\begin{eqnarray}
\omega=2\gamma (C-A\cos p),\;\; A=B<C,\nonumber
\label {ms01}
\end{eqnarray}
corresponding to the linear magnon solution
\begin{eqnarray}
S_n^x=\sqrt{1-\gamma^2}~ \sin(pn-\omega t+\delta) ,\\ \nonumber
S_n^y=\sqrt{1-\gamma^2}~ \cos(pn-\omega t+\delta) , \\ \nonumber
S_n^z=\gamma,
\label {ms02}
\end{eqnarray}
which was noted in ref. [18].

It is also now instructive to analyze the nature of dispersion relations (33)
underlying the nonlinear magnon or elliptic function propagating spin wave 
solutions (34) - (36).

First we note that these solutions (given by (34) - (36)) in the limit 
$k\rightarrow 0$, reduce to the above linear spin wave solutions (40) with 
the specific value $p=\pi$ and $\omega=2\gamma(A+C)$. However, localized 
solitary wave solutions (for $k=1$) do not appear in the moving case because 
$p\rightarrow\infty$ (In fact, in the limit $k\rightarrow 1$, one gets the
ground state solution $(\pm 1, 0, 0)$.  
From the expression (33), we note that for $\gamma=1$, $k=0$ and for 
$\gamma=\gamma_{min}=\sqrt{(B-A)/(B+C)}$ one gets $k=1$.  In other 
words, $1\ge\gamma>\gamma_{min}$.  Defining
$\zeta=(B-A)/(A+C)$ with $B>A$ implies
\be
k^2=\left(\frac{1}{\gamma^2}-1\right)\zeta , ~~~ \gamma^2=\frac{\zeta}
{\zeta+k^2} .
\ee
This leads to the dispersion relation [see Eq. (33)] 
\be
\omega=2\sqrt{\frac{(B-A)(B+C)}{\zeta+k^2}}  ,
\ee
with $p=2K(k)$.  Note that $\pi\le p<\infty$ when $0\le k<1$.  For $k\simeq0$, $K(k)=(\pi/2)(1+k^2/4)$ and we
get the dispersion relation for the magnons (40) as 
\be
\omega=2\sqrt{\frac{(B-A)(B+C)}{(\zeta-4)+4p/\pi}}  ,
\ee
which is finite at $p=\pi$ and zero as $p\rightarrow\infty$.
Similarly, for $k\simeq1$ (but not equal to one), we have $K(k)=ln(4/k')$ and we get the
dispersion for the soliton like structure as 
\be
\omega=2\sqrt{\frac{(B-A)(B+C)}{(1+\zeta)-16\exp(-p)}}  ,
\ee
which is finite at both $p=\pi$ and as $p\rightarrow\infty$. The above 
dispersion relations in the limiting cases of the parameter $k$ in the 
allowed region of $p$, $\pi\le p <\infty$, are plotted in Fig. 3 for a simple
choice of the anisotropy parameters.  Dispersion curves for other values of 
$k$ lie between these two limiting curves.  
 
\begin{figure}[!ht]
\begin{center}
  \includegraphics[width=.5\linewidth]{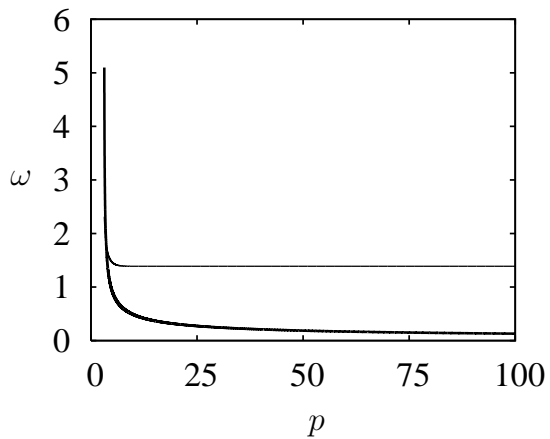}
\vskip 20 pt
\caption{Nonlinear spin wave dispersion relations in the limiting 
cases (i) $k\approx 0$: thick line and (ii) $k \approx 1$: thin line, for
the parameter values $A=2.45, B=2.65$ and $C=0$.  }
\end{center}
\end{figure}

\section{Static solutions} 

Considering Eqs. (18) - (20), one can easily note that the term $[1-k^2 
\sn^2(u,k)\sn^2(p,k)]$ in the denominator on the right hand sides is 
``harmless" provided $\omega=0$ [case(1)] on the left hand sides. This 
actually corresponds to the static case (or time independent case) of the 
equation of motion (6) - (8) for  $D=0$ and $\vec{\cal H}=0$.  In this 
case the conditions (18) - (20) become 
\bea 
\beta\gamma[C\dn(p,k)-B\cn(p,k)]&=&0,\nonumber\\
\gamma\alpha[A\cn(p,k)-C)]& =&0,\nonumber\\ 
\alpha\beta[B-A\dn(p,k)] &= &0 . 
\eea 
Then there are three possibilities. 

\subsection{Nonplanar static structures for XYZ and XYY models}




In this case $S_n^x,S_n^y,S_n^z\ne0$, which implies that $\alpha,\beta, 
\gamma \ne0$.  The static, periodic solutions of such a discrete Heisenberg 
chain are then obtained from (13) and (14) with $u=pn+\delta$ as 

\begin{eqnarray}  
S_n^x=\sqrt{1-\gamma^2 k^{{\pr}2}}~ \sn(pn+\delta,k) ,\\ \nonumber   
S_n^y=\sqrt{1-\gamma^2}~ \cn(pn+\delta,k) , \\ \nonumber   
S_n^z=\gamma~ \dn(pn+\delta,k) ,  
\end{eqnarray}  
with the fixed modulus $k$ and fixed wave vector $p$ given by   
\be 
k^2=\frac{A^2-B^2}{A^2-C^2} , ~~~ \dn(p,k)=\frac{B}{A} . 
\ee 
In (46), $\delta$ is a constant phase factor. Note that the above 
expressions (47) follow from Eq. (45) for $\alpha 
\ne\beta\ne0$.  Here $p$ denotes a wave vector.  
In the special case of an XYY model, i.e. $A\neq B=C$, which implies 
$k=1$ [from Eq. (47)], we get the localized single soliton (kink- and pulse-like) 
solutions or domain wall structures:  
\begin{eqnarray}
S_n^x=\tanh(pn+\delta), ~~~ S_n^y=\sqrt{1-\gamma^2}~\sech(pn+\delta) ~~~ 
S_n^z=\gamma~\sech(pn+\delta) , 
\end{eqnarray}
with $\sech(p)=B/A$.  The above domain wall structure is depicted schematically
in Fig. (4a).  Alternatively, for the XXY model, $A=B\neq C$ (i.e. 
$k=0$), we get the linear excitations (frozen magnons):
\be
S_n^x=\sqrt{1-\gamma^2}~\sin(pn+\delta) , ~~~ S_n^y=\sqrt{1-\gamma^2}~   
\cos(pn+\delta) , ~~~ S_n^z=\gamma . 
\ee 

\subsection{Planar (XY) case} 
There are two other possibilities from Eq. (45) corresponding to the 
planar (XY) case. 

{\bf Case (i)}: In this case $\alpha=\beta=1$ and $\gamma=0$ which 
implies $S_n^z=0$ and $S_n^x,S_n^y\ne0$.  Again from Eqs. (13) and 
(14) the solution is given by 
\be 
S_n^x=\sn(pn+\delta,k) , ~~~ S_n^y=\cn(pn+\delta,k) , 
\ee 
provided $\dn(p,k)=B/A$, where the modulus $k$ ($0\le k\le 1$) and the 
constant $\delta$ are arbitrary, which follows from Eq. (45).  As in 
the XYZ case one can obtain domain wall structure in the XY case also 
by taking the limit $k\rightarrow1$, namely $S_n^x=\tanh(pn+\delta)$ 
and $S_n^y=\sech(pn+\delta)$.  The above solution was already 
reported by Roberts and Thompson \cite{roberts} and Granovskii and 
Zhedanov \cite{granov}.  
 
{\bf Case (ii)}: In this case $\alpha=k$, $\beta=0$ and $\gamma=1$ 
which implies $S_n^y=0$ and $S_n^x,S_n^z\ne0$.  The solution is now 
given by 
\be 
S_n^x=k~\sn(pn+\delta,k) , ~~~ S_n^z=\dn(pn+\delta,k) ,
\ee 
provided $\cn(p,k)=C/A$.  Here also the modulus $0<k<1$ and the 
constant $\delta$ are arbitrary.  The domain structure in the $k 
\rightarrow 1$ limit is again given by $S_n^x=\tanh(pn+\delta)$ and 
$S_n^z=\sech(pn+\delta)$.  This solution was also reported in 
references \cite{roberts} and \cite{granov}.  
 

\subsection{Another class of Nonplanar XYY structures} 

Now making use of the more general parametrization (15), one can 
easily check that in the XYY case, that is $B=C\ne A$ with the 
second variable $v$ fixed as a constant, one can identify the following 
static solution: 
\be 
S_n^x=\cn(pn+\delta,k), ~~~ S_n^y=\gamma~\sn(pn+\delta,k), ~~~ 
S_n^z=\sqrt{1-\gamma^2}~\sn(pn+\delta,k) , 
\ee 
provided $\dn(p,k)=A/B$, while the modulus parameter $k$ ($0\le k\le 1$), 
and the constants $\gamma=\cn(v,k_2)$ and $\delta$ are arbitrary.  In 
the limit $k\rightarrow1$, one obtains the domain wall structure 
\be 
S_n^x=\sech(pn+\delta) , ~~~ S_n^y=\gamma~\tanh(pn+\delta) , ~~~ 
S_n^z=\sqrt{1-\gamma^2}~\tanh(pn+\delta) , 
\ee 
with $\tanh(p)=A/B$. The above type of domain wall structure is 
depicted schematically in Fig. (4b).  It is also of interest to note 
that none of the above static solutions survive in the continuum 
limit of the lattice and they are all patently structures belonging 
to discrete lattices.  

\begin{figure}[!ht]
\begin{center}
  \includegraphics[width=.7\linewidth]{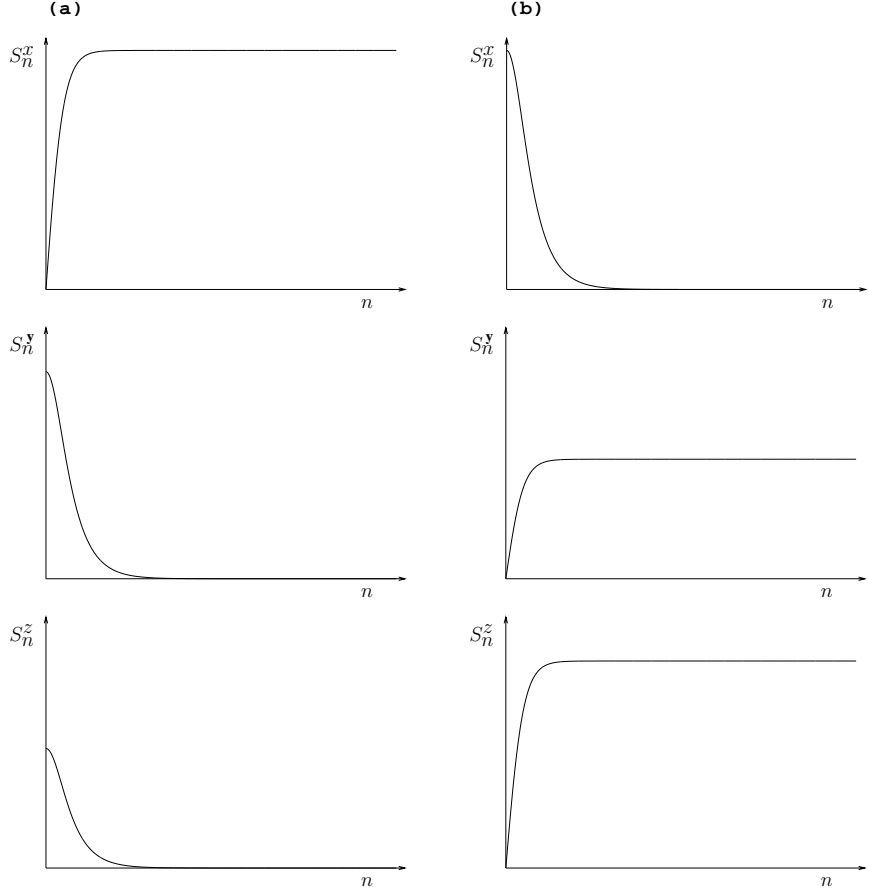}
\vskip 20 pt
\caption{Localized static single soliton/domain wall structures in 
the nonplanar case of XYY model with $\gamma =0.5$: (a) Spin structure given 
by Eq. (48), (b) Spin structure given by Eq. (53).}
\end{center}
\end{figure}

\subsection{Integrability of the static case} 

In their important work \cite{granov}, Granovskii and Zhedanov have 
shown that the static case of Eq. (5), namely 
\be 
\vec{S}_n\times\vec{J}(\vec{S}_{n+1}+\vec{S}_{n-1})=0 , ~~~ 
(\vec{J}\vec{S}_n)=AS_n^x+BS_n^y+CS_n^z , 
\ee 
or equivalently 
\be 
\vec{S}_{n+1}+\vec{S}_{n-1}=\lambda_n J^{-1}\vec{S}_n , 
\ee 
where the Lagrange multiplier 
\be 
\lambda_n=\frac{2(\vec{S}_n\cdot J^{-1}\vec{S}_{n+1})}{\vec{S}_n 
\cdot J^{-2}\vec{S}_n},  
\ee 
is equivalent to a discretized version of the Schr\"odinger equation 
with two-level Bargmann type potential or a discrete analog of a 
Neumann system \cite{veselov}.  Some explicit solutions have also 
been reported in their works.  

\section{Energy values of static structures and Peierls-Nabarro 
potential barrier} 

For each of the static structures discussed in Sec. IV, the total 
energy can be obtained explicitly by making use of the various 
summation formulas given in the Appendix.  In the following we 
indicate how the total energy for the lattice can be obtained for 
the nonplanar static structure (46) and then present the final 
results only for the other cases. 

\subsection{Total energy of the nonplanar static structure (46)} 

For the nonplanar static structure (46), the total energy 
\begin{eqnarray}
E&=&-\sum_{\{n\}}[AS_n^xS_{n+1}^x + BS_n^yS_{n+1}^y + CS_n^zS_{n+1}^z]
\nonumber \\
&=&-\frac{1}{2}\sum_n\{A\alpha^2\sn(u,k)[\sn(u+p,k)+\sn(u-p,k)] 
+B\beta^2\cn(u,k)[\cn(u+p,k)
\nonumber\\&&+\cn(u-p,k)]+ C\gamma^2\dn(u,k)[\dn(u+p,k)+\dn(u-p,k)]\}  .
\end{eqnarray}
Now using the identities derived recently by Khare et al. \cite{khare} 
for the products of elliptic functions, summarized in Appendix A, Eq. 
(57) can be rewritten as 
\begin{eqnarray}
E&=&-\sum_{n=1}^N[B\beta^2\cn(p,k)+C\gamma^2\dn(p,k)]+\sum_{n=1}^N\left( 
-\frac{A\alpha^2}{k^2\sn(p,k)}+\frac{B\beta^2\dn(p,k)}{k^2\sn(p,k)} 
+\frac{C\gamma^2\cn(p,k)}{\sn(p,k)}\right) \nonumber \\ 
&&\times [Z(p(n+1)+\delta,k)-Z(pn+\delta,k)-Z(p,k)] . 
\end{eqnarray}  
Since the elliptic zeta function \cite{byrd} satisfies  
\be 
\sum_{n=1}^N[Z(p(n+1)+\delta,k)-Z(pn+\delta,k)]=0 , 
\ee 
see Ref. \cite{khare}, Eq. (58) reduces to 
\be 
E=-N[B\beta^2\cn(p,k)+C\gamma^2\dn(p,k)]-N\frac{Z(p,k)}{k^2\sn(p,k)} 
[-A\alpha^2+B\beta^2\dn(p,k)+C\gamma^2\cn(p,k)] , 
\ee 
which is independent of the constant phase factor $\delta$.  Consequently, 
the Peierls-Nabarro (P-N) potential barrier vanishes.  Since $\alpha= 
\sqrt{1-\gamma^2k'^2}$, $\beta=\sqrt{1-\gamma^2}$, $k^2=(A^2-B^2)/(A^2-C^2)$ 
and $\dn(p,k)=B/A$, see Eqs. (46) - (47), the energy expression (60) can be 
rewritten as 
\be 
E=-N\frac{BC}{A}-N\frac{\sqrt{A^2-C^2}}{(A^2-B^2)}Z(p,k)\left[-(A^2-B^2) 
+\gamma^2C^2\frac{(B^2-C^2)}{(A^2-C^2)}\right] . 
\ee    
Note that in the above expression $\gamma$ is a free parameter, while all 
the other quantitites are fixed and the energy is a quadratic function of 
the free parameter $\gamma$. 

In general, this energy may depend on the location of the soliton, i.e. 
$\delta$.  There is an energy cost associated with moving a localized mode,
e.g. a soliton or breather, by half a lattice constant in a discrete
lattice.  Alternatively, there is a periodic dependence of the energy of 
a soliton on its position with respect to the lattice sites. This is called 
the Peierls-Nabarro barrier \cite{nabarro,PN}.  The effects of discreteness 
such as the P-N barrier and the spin barrier (i.e., the total spin, which 
is an integral of motion, depends on the location of the soliton) may be 
studied from the total energy expressions \cite{soboleva}. In the present 
case the P-N barrier vanishes as the energy expression is independent of 
the location of the soliton $\delta$.

\subsection{Total energy of other static structures} 

In the following, we only give the final form of the total energy 
expressions for the other static structures. 

{\bf (i) Planar XY case}: 

(a) Structure (50): 
\be 
E=-NB cn(p,k)+N\frac{Z(p,k)}{k^2sn(p,k)}[A-Bdn(p,k)].  
\ee 
Here $\dn(p,k)=B/A$, while the modulus parameter $k$ is arbitrary.  
Consequently the energy expression (62) can be expressed as 
\be 
E=-N\frac{B}{A}\frac{\sqrt{B^2-A^2+A^2k^2}}{k}+N\sqrt{A^2-B^2} 
\frac{Z(p,k)}{k} , 
\ee 
where $k$ ($0\le k \le 1$) is the free parameter. 

(b) Structure (51): 
\be 
E=-NCdn(p,k)+\frac{NZ(p,k)}{k^2sn(p,k)}[Ak^2-Ccn(p,k)]. 
\ee 
Here $\cn(p,k)=C/A$, while $k$ is arbitrary.  Then the expression (64) can 
be written as 
\be 
E=-N\frac{C}{A}\sqrt{A^2-k^2(A^2-C^2)}+N\frac{(A^2k^2-C^2)}{\sqrt{A^2-C^2}} 
\frac{Z(p,k)}{k^2} . 
\ee 
{\bf (ii) Nonplanar XYY structure}: For the solution (52), the total 
energy expression can be deduced as 
\be 
E=-NAcn(p,k)+N\left[Ak^2\frac{dn(p,k)}{sn(p,k)}-[B\gamma^2+C(1-\gamma^2)] 
\frac{1}{k^2sn(p,k)}\right]Z(p,k).
\ee 
In Eq. (66), $\dn(p,k)=A/B$ while $k$ is arbitrary.  Then we have 
\be 
E=-N\frac{A}{B}\frac{\sqrt{A^2-B^2k^2}}{k}+\frac{N}{\sqrt{B^2-A^2}} 
\left[{A^2k^3} -\{C+(B-C)\gamma^2\}\frac{B}{k}\right]Z(p,k) . 
\ee  
Note that in the above equation both $k$ and $\gamma$ are free parameters. 
All the above energy expressions are independent of the phase constant $\delta$
and so the P-N potential barrier in these cases is also absent.
\section{Isotropic model $(A=B=C=1)$}
No moving/time dependent nonlinear structures can be found in this case 
except for the spin wave (magnon) solutions (see below). This can be easily 
checked by looking at the conditions that must hold to satisfy Eqs. (18) - (20)  
which are for $k\neq0$, (for $k=0$, see below)
\be
\omega\alpha=-4\beta\gamma , ~~~ \omega\beta=-4\alpha\gamma, ~~~
\omega\gamma k^2=0 .
\ee
Then the only possible structures are the static structures of the following
form.
\subsection{Planar model: static solutions}
In the special case of an isotropic planar model ($A=B=C$; $S_n^z=0$)
the limiting elliptic function solutions for $\gamma=0$ (and $\omega=0$) are
\be
S_n^x=\sn\left(2Kn+\delta,k\right) , ~~~
S_n^y=\cn\left(2Kn+\delta,k\right) , ~~~ S_n^z=0.
\ee
The modulus parameter $k$ and the phase constant $\delta$ are arbitrary. 
These solutions were obtained previously \cite{roberts}.  However, the 
static solutions analogous to Eq. (50) do not exist in the isotropic case as
$K(k)\rightarrow\infty$ as $k\rightarrow 1$ and the solution (69) reduces to
the uniform solution $S_n=(\pm1,0,0)$.
However, if instead $\beta=0$, the solutions are given by
\be
S_n^x=k~\sn\left(2Kn+\delta,k\right) , ~~~ S_n^y=0 , ~~~
S_n^z=\dn\left(2Kn+\delta,k\right) .
\ee
\subsection{Nonplanar model: propagating solutions}
For $k=0$, $\omega=2\gamma(1-\cos p)=4\gamma \sin^2(\frac{p}{2})$, the propagating
linear excitations (i.e. magnons) are given by
\be
S_n^x=\sqrt{1-\gamma^2}~\sin(pn-\omega t) , ~~~
S_n^y=\sqrt{1-\gamma^2}~\cos(pn-\omega t) , ~~~ S_n^z=\gamma.
\ee

\section{Linear stability} 

\addtocounter{footnote}{+1}
Next, we consider the stability of both the time periodic solutions,
discussed in Sec. III, and the static
solutions in terms of Jacobi elliptic functions, obtained in 
Sec. IV$^1$.\footnotetext{The work in this section was carried out
in collaboration with S. Murugesh}
Linear stability of the time periodic solutions in Eqs. $(24-26)$ and 
Eqs. $(34-36)$ is studied using the period map. For the homogeneous time
periodic solution, it suffices to study the stability of 
spin at any one site. For the spatially oscillatory solution in Eqs. $(34-36)$,
we individually study the stability of two spin vectors, one at an odd and 
another at an even site. Rewriting the spin equation
$(5)$ using the complex stereographic variable
\be
\Omega_n = \frac{S^x_n + iS^y_n}{1+S^z_n},
\ee
we get
\bea
\frac{d\Omega_n}{dt} = iC\Omega_n\Big{(}
\frac{1-|\Omega_{n+1}|^2}{1+|\Omega_{n+1}|^2} +
\frac{1-|\Omega_{n-1}|^2}{1+|\Omega_{n-1}|^2} \Big{)}
-i\frac{A}{2}\Omega_n^2\Big{(} 
\frac{\Omega_{n+1} + \bar{\Omega}_{n+1}}{1+|\Omega_{n+1}|^2} +
\frac{\Omega_{n-1} + \bar{\Omega}_{n-1}}{1+|\Omega_{n-1}|^2} \Big{)}\\\nonumber
+i\frac{B}{2}\Omega_n^2\Big{(} 
\frac{\Omega_{n+1} - \bar{\Omega}_{n+1}}{1+|\Omega_{n+1}|^2} +
\frac{\Omega_{n-1} - \bar{\Omega}_{n-1}}{1+|\Omega_{n-1}|^2} \Big{)}
+ 2iD\frac{1-|\Omega_n|^2}{1+|\Omega_n|^2}\Omega_n
+ i\frac{H_x}{2}(1+\Omega^2_n) , 
\eea
where $\bar{\Omega}_n$ denotes the complex conjugate of the stereographic 
variable. After linearizing using the expansion
\be
\Omega_n = \Omega_{0n} +\delta \Omega,
\ee
around the time periodic solution $\Omega_{0n}(t)$,
we compute the Floquet matrix $\hat{\mathcal{M}}$ such that
\be
\left(\begin{array}{c}
\delta{\Omega}_n(T)\\
\delta{\bar{\Omega}}_n(T)
\end{array}
\right)
 = \hat{\mathcal{M}}
\left(\begin{array}{c}
\delta\Omega_n(0)\\
\delta\bar{\Omega}_n(0)
\end{array}
\right).
\ee
Here, $T=2\pi/\omega$ is the inherent time period in the two sets of solutions,
Eqs. (24) - (26) and (34) - (36). If $\gamma_n$ is the eigenvalue of 
$\mathcal{M}$, then the solution is unstable if $|\gamma_n|>1$. Figure 5 shows 
the instability regions in the $(k-A)$ plane for the homogeneous and spatially 
oscillatory time periodic solutions. Here $k$ is the modulus of the Jacobian 
elliptic function. 

\begin{figure}[h]
\begin{center}
\includegraphics[width=1.2\linewidth]{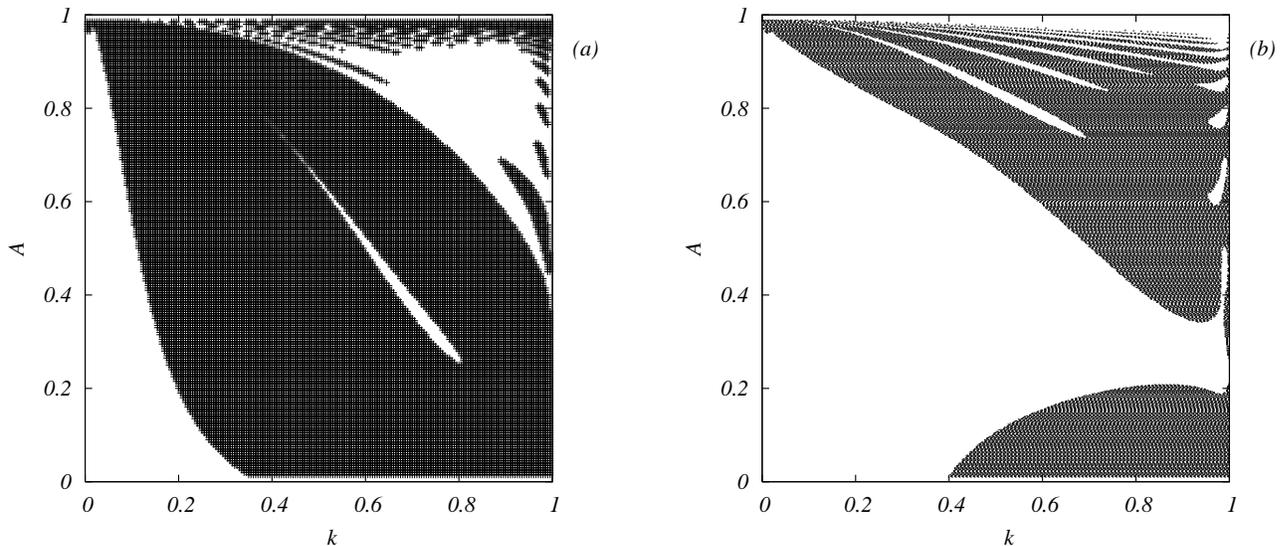}
\caption{Instability regions (shaded regions) in the $(k-A)$ plane for the time 
periodic solutions obtained using the period map. Instability region for the 
(a) homogeneous time periodic solution, Eqs. (24) - (26), with $B= 0.9$ 
and $C=0.01$, $B>A>C$, and (b) for the spatially oscillatory solution, Eqs. 
(34) - (36), for the same parametric values. The instability diagram is 
identical for spins at both odd and even sites. }
\end{center}
\end{figure}

In order to numerically study the stability of the static solutions of Sec. IV,
we perturb the solution ${\bf S}^0_n$ in Eqs. (46) - (47) by a small amount 
$\delta {\bf S}_n(t)$
such that ${\bf S}_n\cdot\delta{\bf S}_n = 0$, $|\delta{\bf S}_n|\ll1$.
Upon substituting the perturbed vector 
\be
{\bf S}^p_n = {\bf S}^{0}_n + \delta{\bf S}_n
\ee
in Eq. $(5)$, the time evolution is computed numerically. As an illustration, 
it is found that the static solution (46) - (47) is indeed stable for long 
times for small values of the modulus parameter $k$ of the Jacobi elliptic 
function, and that the solution (46) - (47) is less stable with increasing 
value of $k$, i.e. the instability sets in at earlier times. Figure 6 shows 
the time profile of the static solution (46) - (47) under a small perturbation. 
Figure 7 depicts the initial and final profiles for easy comparison.  Fuller 
details will be presented elsewhere. 

Finally, it is also of interest to investigate whether the time periodic 
solutions (24) - (26) and (34) - (36) are modulationally stable or not.  
Recently such modulational stability analysis has been performed for special 
solutions of a number of discrete nonlinear dynamical systems, including 
discrete nonlinear Schr\"odinger equations, see for example refs. 
\cite{rapti1,rapti2}.  Such an analysis for the time dependent elliptic 
function solutions (24) - (26) and (34) - (36) is being pursued at present 
and will be reported separately along with the details of the linear stability 
analysis mentioned above.  
 
\begin{figure}[h]
\begin{center}
\includegraphics[width=.7\linewidth]{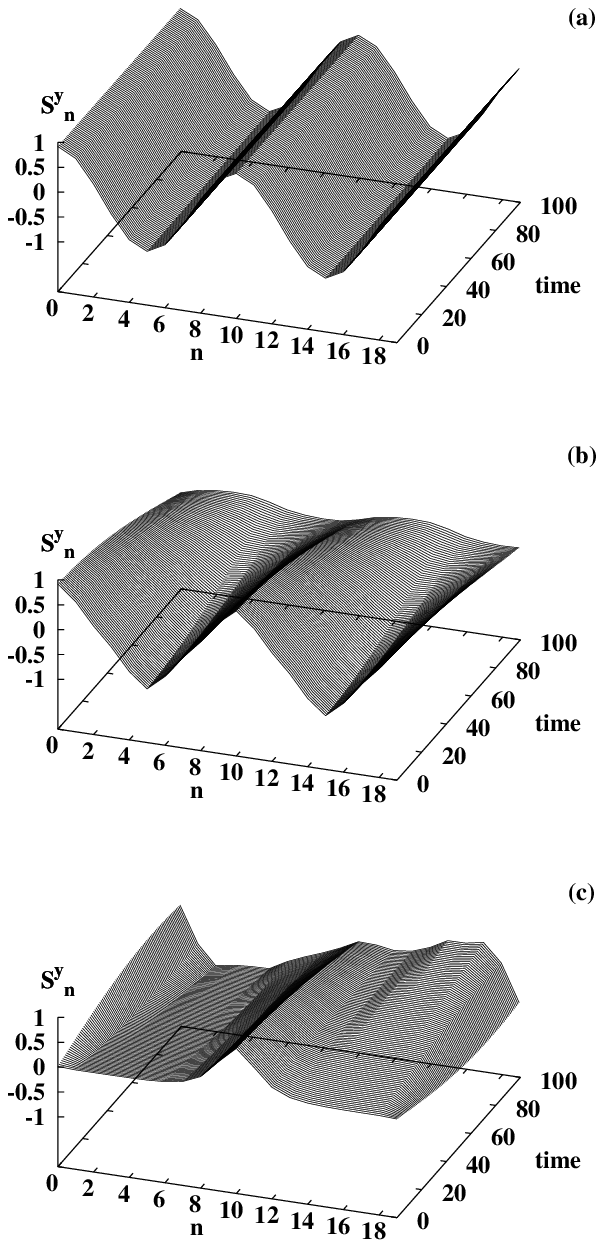}
\caption{Time evolution of the $S_n^y$ component of the static solution, 
Eqs. (46) - (47), under
perturbation, (a) $k=0.3$, (b) $k=0.9$ and (c) $k=1$. As can be noticed, for 
small
values of $k$, the solution tends to be more stable for long periods of time.}
\end{center}
\end{figure}

\begin{figure}[h]
\begin{center}
\includegraphics[width=.6\linewidth]{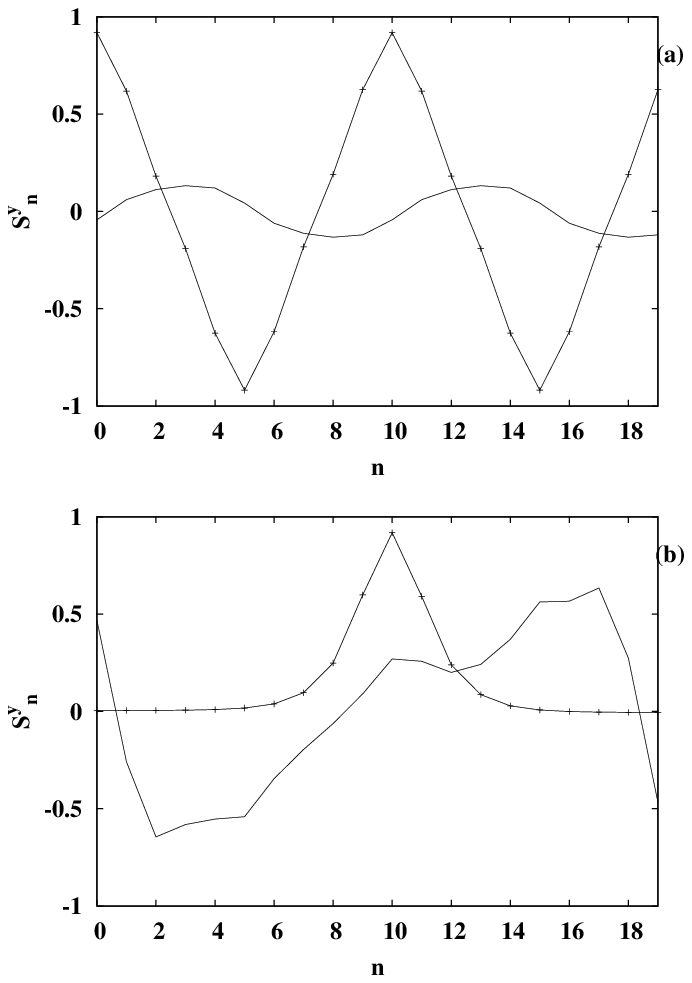}
\caption{Initial (line-points) and final (line) profiles of the static solution 
(46) - (47) under perturbation (Eq. $(76)$) in (a) Figure 6(b) with $k=0.9$ and 
(b) Figure 6(c) with $k=1$. Instabilities start to appear at earlier time 
as $k$ approaches $1$. }
\end{center}
\end{figure}

\section{Anisotropic spin chain in the presence of on-site anisotropy and a 
constant external magnetic field}
Finally, in this section we wish to point out that explicit spin solutions can
be constructed for even more general situations (i) with additional on-site
anisotropy and (ii) external constant magnetic field.  Brief details are as
follows.
\subsection{On-site anisotropy}
Many of the results discussed in the previous sections are valid even in the
presence of the on-site anisotropy ($D\neq0$) in Eq.(5) or Eqs. (6) - (8).  
In this case, the parametrization (17) in terms of Jacobian elliptic functions 
leads to the following conditions instead of (18) - (20):
\begin{mathletters}
\begin{eqnarray} 
-\omega\alpha=2\beta\gamma \bigg\{\frac{[C\dn(p,k)-B\cn(p,k)]}{1-k^2\sn^2(u,k) 
\sn^2(p,k)}-D \bigg\} , \\
\omega\beta=2\alpha\gamma \bigg\{\frac{\dn(p,k)[A\cn(p,k)-C]}{1-k^2\sn^2(u,k)   
\sn^2(p,k)}+D \bigg\} ,  \\
\omega\gamma k^2=\frac{2\alpha\beta\cn(p,k)[B-A\dn(p,k)]}{1-k^2 
\sn^2(u,k)\sn^2(p,k)} ,  ~~~ u=pn-\omega t + \delta .    
\end{eqnarray}
\end{mathletters}

Unlike the $D=0$ case (sec. III A), the above  relations can be free
from the space-time variable ($u$) only for two choices: (i) modulus parameter
$k=0$ (linear spin wave solution) and (ii) $p=4K(k)$ (spatially homogeneous
time-dependent solution) or $p=2K(k)$ (spatially oscillatory time periodic
solution). 
\begin{itemize}
\item[(i)   ] For the case $k=0$, from Eq. (77c), we have $B=A$ and from 
Eqs. (77a), (77b) we obtain the  dispersion relation 
\begin{equation}
\omega = 2 \gamma (C-D-A \mbox{cos}~ p), \;\; A=B< (C-D),
\end{equation}
associated with the  spin wave solution (40) for the present case where the
on-site anisotropy $D \neq 0$.

\item[(ii-a)] For the choice $p=4K(k)$, Eqs. (77) degenerate to 
\begin{equation}
\omega = 2 \gamma \sqrt{(C-B-D)(A-C+D)}, \;\; k^2 =
\frac{(1-\gamma^2)}{\gamma^2} \frac{(B-A)}{(B+D-C)},
\end{equation}
where the form of the associated spatially homogeneous time-dependent solutions
are given by Eqs. (24) - (26) for the spin components $S_n^x$, $S_n^y$ and
$S_n^z$ but with the above expressions for $\omega$ and $k^2$.

\item[(ii-b)] Similarly for the choice $p=2K(k)$, Eqs. (77) lead to the expressions
\begin{equation}
\omega = 2 \gamma \sqrt{(A+C-D)(B+C-D)}, \;\; k^2 =
\frac{(1-\gamma^2)}{\gamma^2} \frac{(B-A)}{(A+C-D)}
\end{equation}
for the spatially oscillatory time periodic solutions (34) - (36) with the above
expressions for $\omega$ and $k^2$.
\end{itemize}

It may be noted from the  relation (77) that no static solution (that is
$\omega =0$) is possible for general values of $k$ when $D\neq 0$.

\subsection{Constant external magnetic field}
Considering now the full anisotropic chain (6) - (8) with the external magnetic
field also present, we have not succeeded in finding any explicit solution
which generalizes the nonlinear structures 
discussed in the previous sections for the anisotropic spin chain in 
the absence of the magnetic field.  However, in specific instances classes of
exact solutions can be obtained.  For example, for the XYY spin chain, with 
$B=C\ne A$, $D=0$,  Eq. (5) or Eqs. (6) - (8) admit(s) the following 
exact solutions.
In the presence of an external magnetic field $\vec{H}=(H_x,0,0)$ 
the relevant contribution to the equation of motion for the anisotropic 
spin chain comes from $\vec{S}_n\times\vec{H}$:   
\be \frac{d{S_n^x}}{dt}=C[S_n^y(S_{n+1}^z+S_{n-1}^z)
-S_n^z(S_{n+1}^y+S_{n-1}^y)] ,
\ee
\be \frac{d{S_n^y}}{dt}=AS_n^z(S_{n+1}^x+S_{n-1}^x)
-CS_n^x(S_{n+1}^z+S_{n-1}^z) + H_xS_n^z,
\ee
\be \frac{d{S_n^z}}{dt}=CS_n^x(S_{n+1}^y+S_{n-1}^y)
-AS_n^y(S_{n+1}^x+S_{n-1}^x) - H_xS_n^y.
\ee
These equations have an exact solution with $\dn(p,k)=C/A$ 
and $\omega=H_x$ in the form 
\begin{eqnarray} 
S_n^x=\sn(pn+\delta,k), \nonumber \\ 
S_n^y=\sin(\omega t+\gamma) \cn(pn+\delta,k), \nonumber \\ 
 S_n^z=\cos(\omega t+\gamma)\cn(pn+\delta,k) . 
\end{eqnarray}  
In Eq. (84), the modulus parameter $k$ ($0\le k \le 1$) is arbitrary, while 
$\gamma$ and $\delta$ are arbitrary phase factors. 
These solutions can be inferred by generalizing the static spin structures of
the $XYY$ model discussed in Sec. IV. C.  Similar solutions can also be written
down when the  magnetic field is along the $y$ or $z$ direction. 

For the present case the total energy becomes
\begin{eqnarray}
E &=& -A \sum_n S_n^x S_{n+1}^x-C \sum_n(S_n^y S_{n+1}^y+S_n^z S_{n+1}^z) -H_x \sum_n
S_n^x \nonumber \\
&=& -A \sum_n \mbox{sn} (pn+\delta)\;\mbox{sn}(p(n+1)+\delta) \nonumber \\
& & \hspace{1cm}-C \sum_n
\mbox{cn}(pn+\delta)\; \mbox{cn}(p(n+1)+\delta) -H \sum_n \mbox{sn}(pn+\delta)
\nonumber \\
&=& -N C \;\mbox{cn}(p,k)-\frac {N Z(p,k)}{k^2\; \mbox{sn}(p,k)} [-A+C \;\mbox{dn}
(p,k)] -H_x \sum_n \mbox{sn} (pn+\delta,k) \nonumber \\ 
&=& -N\frac{C}{A}\frac{\sqrt{C^2-A^2+k^2A^2}}{k} + N\frac{\sqrt{A^2-C^2}}{k} 
Z(p,k) + H_x \sum_n \mbox{sn} (pn+\delta,k) . 
\end{eqnarray}
The sum in the last term above, namely $\sum_n \mbox{sn}(pn+\delta)$, is
represented as $\sigma_3(\delta)$ in Ref. \cite{khare} by Khare and Sukhatme 
and is dependent on the location of the soliton $\delta$.  This ensures that the
Peierls-Nabarro potential barrier is present in the anisotropic spin chain in 
the presence of an external magnetic field.
\section{Summary and Conclusions} 
Using the summation identities \cite{khare} for Jacobi elliptic functions 
\cite{byrd} we obtained several classes of static and propagating 
exact solutions for the classical, anisotropic Heisenberg chain. 
In the special case of the isotropic planar model we recovered 
the previously known solutions \cite{roberts}.  We explicitly 
obtained the nontrivial dispersion relations ($\omega$ vs. $p$) 
for the propagating solutions and predicted the contrasting 
features of magnons and solitons.  Specifically, as $p\rightarrow 
\infty$ the magnon frequency goes to zero whereas the soliton 
frequency reaches a nonzero value.  These dispersion relations 
can be measured via neutron scattering in the quasi-one dimensional 
materials realized by anisotropic Heisenberg chains.  It would be 
instructive to explore whether similar exact solutions can be 
obtained for the analogous quantum Heisenberg models. The effects 
of discreteness, e.g. Peierls-Nabarro barrier \cite{nabarro,PN,soboleva} 
and spin barrier \cite{soboleva}, may be important in anisotropic 
spin chains. 
The discrete equation of motion is non-integrable in general.  
However, the static version in the absence of an external field 
is an integrable system \cite{discrete2}. 
It is instructive to explore semiclassical quantization of the $2K$
versus $4K$ periodic solutions in terms of $N$ anharmonic oscillators
\cite{elstner}.
The solutions expressed in terms of $\sn(x,k)$, $\cn(x,k)$ and $\dn(x,k)$
correspond to the $n=1$ Lam\'e functions.  We have explicitly
checked that $n=2$ Lam\'e functions do not give exact solutions.
However, it is conceivable that $n=3$ Lam\'e functions \cite{lame}
may lead to a new class of exact solutions.  In addition, there
may be another class of exact solutions with a denominator also
containing elliptic functions. We are presently exploring these 
solutions also.  To conclude, we wish to state that the classical 
anisotropic Heisenberg spin chain admits very interesting static 
and dynamic structures and more work is needed in this direction 
to identify all of them.

\section{Acknowledgments} 
M.L. acknowledges the hospitality of the Center for Nonlinear
studies at LANL.  This work was supported in part by the U.S.
Department of Energy. The work of ML forms part of a Department of Science and 
Technology, Government of India sponsored research project and is supported by a
DST Ramanna Fellowship.

\appendix
\section{JACOBIAN ELLIPTIC FUNCTIONS} 

{\it Basic elliptic function properties:}

\be 
\sn^2(u,k)+\cn^2(u,k)=1 , ~~~ \dn^2(u,k)+k^2\sn^2(u,k)=1 . 
\ee 

{\it Addition theorems:}
\be 
\sn(u+v,k)+\sn(u-v,k)=\frac{2\sn(u,k)\cn(v,k)\dn(v,k)}{1-k^2\sn^2(u,k) 
\sn^2(v,k)} , 
\ee 
\be 
\cn(u+v,k)+\cn(u-v,k)=\frac{2\cn(u,k)\cn(v,k)}{1-k^2\sn^2(u,k)  
\sn^2(v,k)} , 
\ee  
\be 
\dn(u+v,k)+\dn(u-v,k)=\frac{2\dn(u,k)\dn(v,k)}{1-k^2\sn^2(u,k)  
\sn^2(v,k)} . 
\ee 
 
{\it Product relations:}
\be
m\sn(x,k)\sn(x+a,k) = -{\rm ns}(a,k)[Z(x+a) - Z(x) - Z(a)],
\ee
\be
m\cn(x,k)\cn(x+a,k) = m\cn(a,k)+ {\rm ds}(a,k)[Z(x+a) - Z(x) -Z(a)], 
\ee
\be
\dn(x,k)\dn(x+a,k) = \dn(a,k) + {\rm cs}(a,k)[Z(x+a) - Z(x) -Z(a)],
\ee
where $Z(x)=Z(x,k)$ is the Jacobi or elliptic zeta function, and 
${\rm ns}(x,k)=1/\sn(x,k)$, ${\rm ds}(x,k)=\dn(x,k)/\sn(x,k)$, 
${\rm cs}(x,k)=\cn(x,k)/\sn(x,k)$.

{\it Summation relation:}
\be
\sum^N_{n=1}\big{\{}Z[\beta\epsilon(n+1)+\delta,k] - 
Z[n\beta\epsilon + \delta,k]\big{\}} = 0.
\ee

{\it Integration formula:}
\be
\int^K_0 \frac{{\sn}^2u du}{1-\alpha^2{\sn}^2 u} = 
\frac{1}{\alpha^2}[\Pi(\alpha^2,k) - K(k)],
\ee
where $K(k)$ and $\Pi(\alpha^2,k)$ are the complete elliptic integrals
of the first and third kind, respectively. 

\section{SEMICLASSICAL QUANTIZATION}
For the spatially homogeneous time-dependent solution (24) - (26), the 
canonically conjugate variables $q_i$ and $p_i$ given by Eq. (30) can be 
expressed in terms of elliptic functions as
\be
p_i = S^z_i = \gamma{\rm dn}u~;~~ u = \omega t+\delta,
\ee 
\be
q_i = \arctan\left(\frac{S_i^y}{S_i^x}\right) = 
\arctan\left(-\sqrt{\frac{1-\gamma^2}{1-\gamma^2k^{'2}}}\frac{\cn u}{\sn u}\right).
\ee
Then the left hand side of the semiclassical quantization condition  (29)
becomes \cite{byrd}
\be
\oint p_idq_i 
= \gamma\sqrt{1-\gamma^2}\sqrt{1-\gamma^2k^{'2}}\int_0^{4K(k)}\frac{{\dn}^2u~du}{1-\gamma^2{\dn}^2u}
\ee
\be
= \frac{4}{\gamma}\sqrt{\frac{1-\gamma^2k^{'2}}{1-\gamma^2}}
\bigg{[}\Pi\bigg(\frac{-\gamma^2k^2}{(1-\gamma^2)},k \bigg)-(1-\gamma^2)K(k)\bigg{]},
\ee
where $K(k)$ and $\Pi(\nu,k)$ are the complete elliptic integrals of the
first and third kind, respectively.

\end{document}